\documentclass{ws-procs9x6}
\def\theequation{\thesection.\arabic{equation}}
\newcommand{\dd}{{\rm d}}

\newcommand{\cN}{\mathcal{N}}
\newcommand{\cnt}{\cN=2}
\newcommand{\cno}{\cN=1}
\newcommand{\cG}{\mathcal{G}}
\newcommand{\cF}{\mathcal{F}}
\newcommand{\cL}{\mathcal{L}}
\newcommand{\sPP}{\mathrm{I}\kern -1.6pt \mathrm{P}}
\newcommand{\sR}{\mathrm{I}\kern -1.6pt \mathrm{R}}

\newcommand{\1}{1\kern -3pt \mathrm{l}}
\newcommand{\tr}{{\rm tr}}

\newcounter{abc}
\def\vev#1{ \langle {#1} \rangle }

\newdimen\tableauside\tableauside=1.0ex
\newdimen\tableaurule\tableaurule=0.4pt
\newdimen\tableaustep
\def\phantomhrule#1{\hbox{\vbox to0pt{\hrule height\tableaurule
width#1\vss}}}
\def\phantomvrule#1{\vbox{\hbox to0pt{\vrule width\tableaurule
height#1\hss}}}
\def\sqr{\vbox{%
  \phantomhrule\tableaustep
\hbox{\phantomvrule\tableaustep\kern\tableaustep\phantomvrule\tableaustep}%
  \hbox{\vbox{\phantomhrule\tableauside}\kern-\tableaurule}}}
\def\squares#1{\hbox{\count0=#1\noindent\loop\sqr
  \advance\count0 by-1 \ifnum\count0>0\repeat}}
\def\tableau#1{\vcenter{\offinterlineskip
  \tableaustep=\tableauside\advance\tableaustep by-\tableaurule
  \kern\normallineskip\hbox
    {\kern\normallineskip\vbox
      {\gettableau#1 0 }%
     \kern\normallineskip\kern\tableaurule}%
  \kern\normallineskip\kern\tableaurule}}
\def\gettableau#1 {\ifnum#1=0\let\next=\null\else
  \squares{#1}\let\next=\gettableau\fi\next}
\newcommand{\Yfund}{\tableau{1}}
\newcommand{\Ysymm}{\tableau{2}}
\newcommand{\Yasymm}{\tableau{1 1}}

\newcommand{\beq}{\begin{equation}}
\newcommand{\eeq}{\end{equation}}
\newcommand{\bea}{\begin{eqnarray}}
\newcommand{\eea}{\end{eqnarray}}

\def\draftnote{}

\begin{document}

\title{
\vspace{-5\baselineskip}
\begingroup
\footnotesize\normalfont\raggedleft
\lowercase{\sf hep-th/0407179} \\
BRX-TH-545\\
BOW-PH-132\\
\vspace{\baselineskip}
\endgroup
Many roads lead to \\$\cnt$ Seiberg--Witten theory\footnote{
\uppercase{B}ased on the talk by \uppercase{H}.\
\uppercase{S}chnitzer at the  \uppercase{D}eserfest,
\uppercase{U}niversity of  \uppercase{M}ichigan, 3--5
\uppercase{A}pril, 2004 } }

\author{STEPHEN G.~NACULICH\footnote{\uppercase{R}esearch
supported in part by the \uppercase{NSF} under grant 
\uppercase{PHY}-0140281.}}

\address{Department of Physics\\
Bowdoin College\\
Brunswick, ME 04011 USA}

\author{HOWARD J.~SCHNITZER\footnote{\uppercase{R}esearch
supported in part by the \uppercase{DOE} under grant
\uppercase{DE--FG02--92ER40706}.} }

\address{Martin Fisher School of Physics\\
Brandeis University\\
Waltham, MA 02454 USA}

\maketitle

\abstracts{
The Seiberg--Witten solution plays a central role 
in the study of $\cnt$ supersymmetric gauge theories.  
As such, it provides a proving ground for a wide variety 
of techniques to treat such problems.  In this review
we concentrate on the role of IIA string theory/M theory 
and the Dijkgraaf--Vafa matrix model, though integrable models 
and microscopic instanton calculations are also of 
considerable importance in this subject.
}

\noindent{\bf Outline}

\begin{itemize}
\item[I.] Introduction
\item[II.] Review of $\cnt$ Seiberg--Witten theory
\item[III.] The SW curve from IIA string theory/M theory
\item[IV.] The Dijkgraaf--Vafa matrix model approach
\item[V.] Concluding remarks
\end{itemize}

\section{Introduction}

Neither of the authors has had the privilege of collaborating
with Stanley Deser, much to our regret.  Since Stanley is still in
the full bloom of his career, there is still a great deal of time
to remedy this.  Stanley has been a wonderful colleague, with whom
we have enjoyed talking about physics and many other topics for
many, many years.  We hope this situation will continue well into
the future.  We dedicate this article to a celebration of
Stanley's achievements.

The theme of our discussion is to argue that $\cnt$
Seiberg--Witten (SW) theory\cite{001}  lies at the cross-roads of
many different aspects of gauge theories.  The various paths that
intersect SW theory are quantum field theory, string theory,
integrable models, and most recently the Dijkgraaf--Vafa (DV)
matrix model\cite{002}.  Because of time limitations, we will not
survey all these roads, but concentrate on just two because ~1)
they are of current interest and ~2) they seem to involve
drastically different concepts, which nevertheless lead to a
better understanding of SW theory.  These different approaches to
a common problem enhance our confidence in the various methods,
and lends support to applying these techniques to new problems.

\begin{picture}(320,330)(0,100)
\put(0,360){\framebox(60,50){{\shortstack{Integrable \\ systems }}}}
\put(60,360){\vector(1,-1){20}}
\put(120,360){\framebox(60,50){{\shortstack{Microscopic \\ instanton \\
calculations}}}}
\put(150,360){\vector(0,-1){20}}
\put(180,385){\vector(1,0){60}}
\put(240,385){\vector(-1,0){60}}
\put(240,360){\framebox(60,50){{\shortstack{Nekrasov's \\ approach}}}}
\put(240,360){\vector(-1,-1){20}}
\put(80,290){\framebox(140,50){{\shortstack{ $\cnt$ SW theory}}}}
\put(0,220){\framebox(100,50){{\shortstack{M-theory/ \\ IIA string theory }}}}
\put(100,270){\vector(1,1){20}}
\put(200,220){\framebox(100,50){{\shortstack{Dijkgraaf-Vafa \\ matrix model }}}}
\put(200,270){\vector(-1,1){20}}
\put(200,150){\framebox(100,50){{\shortstack{Statistical \\ matrix models }}}}
\put(250,200){\vector(0,1){20}}
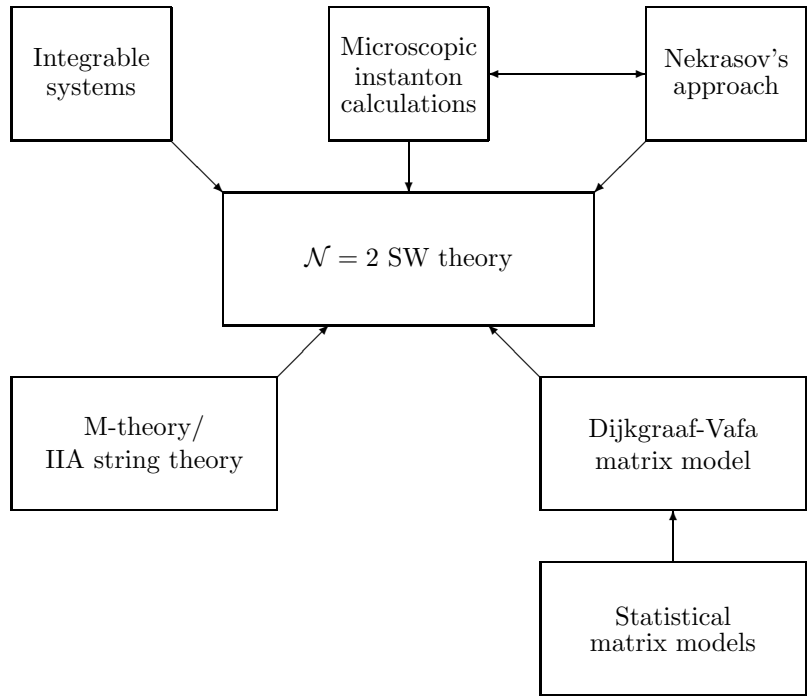
\put(0,120) {\makebox(300,20){{\bf Figure 1:  Roads to Seiberg-Witten theory}}}
\end{picture}

In our own endeavors on SW theory we have been fortunate to have
had a number of excellent collaborators:  Isabel Ennes, Marta
G\'omez-Reino, Carlos Lozano, Henric Rhedin, and Niclas Wyllard.

\section{Review of $\cnt$ Seiberg--Witten theory}
\renewcommand{\theequation}{2.\arabic{equation}}
\setcounter{equation}{0}

$\cnt$ supersymmetric (susy) Yang--Mills theory with gauge group
$\cG$ and hypermultiplets in representation $R$ can be described
in the low-energy region exactly by the SW effective theory.  The
underlying microscopic Lagrangian is
\bea
\label{eq:micro}
{\cL}_{\rm micro}  =  
&-& \frac{1}{4g^2} \: F^a_{\mu\nu} \, F^{\mu\nu a}
+ \frac{\theta}{32\pi^2} \: F^a_{\mu\nu} \, \tilde{F}^{\mu\nu a} \nonumber \\
& +&  \frac{1}{g^2}\: \tr D_\mu \, \phi \, D^\mu \phi^\dagger
-  \frac{1}{2 g^2} \: \tr [\phi ,\phi^\dagger ]^2
\eea
plus fermion and hypermultiplet terms.  
In eq.~(\ref{eq:micro}),
$F^a_{\mu\nu}$ and $\phi^a$ denote the gauge field strength and the bosonic
components of the $\cno$ chiral superfield belonging to the
$\cnt$ vector multiplet respectively, 
both in the adjoint representation. 
The vacuum is defined by
 \beq
 [\phi ,\phi^\dagger ]=0
 \eeq
which implies that $\phi$ may be diagonalized to give
 \beq
\phi = {\rm diag}\: (a_i)\;.
 \eeq
Generically this breaks ${\cG}$ to
 U(1)$^{{\rm rank}\:{\cG}}$.  If only $\phi$ acquires a vacuum
 expectation value (vev), this defines the Coulomb branch.  If
 only the scalar fields in the matter hypermultiplet have vevs,
 this defines the Higgs branch.  We focus on the Coulomb branch.

 The Seiberg--Witten program for $\cnt$ susy gauge
 theories centers on the low-energy effective Lagrangian\cite{001}
 \beq
 {\cL}_{\rm eff}  =  \frac{1}{4\pi} \; {\rm Im} \left[ \int d^4\theta \:
 \frac{\partial {\cF} (A)}{\partial A^i} \: \bar{A}^i  
+  \frac{1}{2} \int d^2\theta \: \frac{\partial^2 {\cF}
(A)}{\partial A^i\partial A^j} \; W^iW^j \right] 
 \eeq
 where ${\cF}(A)$ is the holomorphic prepotential and $A^i$ are
$\cno$ chiral superfields.  
On the Coulomb branch one may write
 \bea
 {\cL}_{\rm eff}  = \frac{1}{4\pi} \; 
\bigg[&-&\frac{1}{4} \, {\rm Im} (\tau_{ij}) F_{\mu\nu}~\!\!^i \, F^{\mu\nu j}
+  \frac{1}{4} \, {\rm Re} (\tau_{ij}) F_{\mu\nu}~\!\!^i \, \tilde{F}^{\mu\nu j}
\nonumber
\\
  & +&  {\rm Im} \,  \partial_\mu (a^\dagger)^j \, \partial^\mu (a_D)_j 
+ \: {\rm fermions} \: \bigg]
  \eea
  where

\renewcommand{\theequation}{\arabic{section}.\arabic{equation}\alph{abc}}
\setcounter{abc}{1}
\bea
a_i &= &{\rm order \; parameters} \\ [8pt]
\addtocounter{equation}{-1}\addtocounter{abc}{1}
  (a_D)_j =  \frac{\partial{\cF}(a)}{\partial a_j}
  &=&  {\rm dual \; order \; parameters}\\
\addtocounter{equation}{-1}\addtocounter{abc}{1}
\label{eq:taudef}
  \tau_{ij} =  \frac{{\partial}^2{\cF}(a)}{\partial a_i\partial a_j} 
    &=& {\rm coupling \; matrix \; or \; period \; matrix.} 
\eea
\renewcommand{\theequation}{\arabic{section}.\arabic{equation}}
One requires Im $\tau_{ij} \geq 0$ for a positive kinetic
energy.

As a result of non-renormalization theorems, the holomorphic
prepotential only receives perturbative corrections at one-loop,
but there is an infinite series of instanton contributions.  Thus
 \beq
 {\cF}(A)  =  {\cF}_{\rm classical} +  {\cF}_{1-{\rm loop}}
  +  \sum^\infty_{d=1} \: \Lambda^{bd} \:  {\cF}_{d{\rm -instanton}}
 \eeq
where
 \bea
 \Lambda & = & {\rm the \; quantum \; scale \; (Wilson \; cutoff)},
 \nonumber \\
 b & = & {\rm the \; coefficient \; of \; the}\; \beta {\rm~function},
 \eea
with $b$ positive for asymptotically-free theories.  For example,
for $\cnt$ SU($N$) theory with $N_f$ hypermultiplets in the
fundamental representation
 \bea
 {\cF}_{\rm 1-loop} (a) & = & \frac{i}{8\pi} \sum^{N}_{i=1}
 \sum_{j\neq i} (a_i - a_j)^2 \log \left(\frac{a_i
 -a_j}{\Lambda}\right)^2 \nonumber \\[.1in]
 && - ~ \frac{i}{8\pi} \, \sum^{N}_{i=1}
 \sum^{N_f}_{I=1} (a_i + m_I)^2 \log \left(\frac{a_i
 + m_I}{\Lambda}\right)^2
 \eea
 yielding 
 \beq
 \tau_{ij} (a) \sim \log (a_i -a_j ) + \ldots  \; \qquad (i \neq j)
\eeq
 at large $a$.  Significantly $\tau_{ij}$ is not single-valued, and
a Riemann surface emerges
for which $\tau_{ij}$ is the period matrix.

In general,
the Seiberg--Witten data needed to find the prepotential are:
\begin{itemize}
 \item[1)] A Riemann surface or algebraic curve specific
to the gauge group and matter content,
dependent on moduli $u_i$, 
which are related to the order parameters $a_i$ ;
 \item[2)] a preferred meromorphic 1-form $\lambda$ = the SW differential;
 \item[3)] a canonical basis of homology cycles on the surface $(A_k, B_k)$.
\end{itemize}

Given this data, the program to solve for the prepotential is:
\begin{itemize}
 \item[1)] 
 Compute the period integrals of the SW differential
\beq
2\pi i \: a_k = \oint_{A_k} \lambda, \qquad \qquad 
2\pi i \: (a_D)_k = \oint_{B_k} \lambda \; . 
\eeq
 \item[2)] Integrate $a_D$ to find ${\cF}(a)$.
 \item[3)] Test this against explicit results from ${\cL}_{\rm micro}$ where possible.
 \item[4)] Test this against the predictions of the DV matrix model.
 \end{itemize}

\begin{picture}(320,300)(0,0)
\put(0,240){\framebox(100,50){{\shortstack{Microscopic \\ Lagrangian\\
					for $\cnt$ susy YM }}}}
\put(110,240){\framebox(100,50){{\shortstack{M-theory or \\ 
                                            geometric engineering}}}}
\put(22,240){\vector(0,-1){20}}
\put(77,240){\vector(0,-1){20}}
\put(160,240){\vector(0,-1){20}}
\put(0,170){\framebox(45,50){{\shortstack{ Instanton \\ calculus}}}}
\put(55,170){\framebox(45,50){{\shortstack{ Nekrasov \\ approach}}}}
\put(110,170){\framebox(100,50){{\shortstack{ SW curve and \\
differential for \\ Coulomb branch \\
of $\cnt$ susy YM}}}}
\put(240,170){\framebox(60,50){{\shortstack{DV\\
matrix \\model }}}}
\put(240,195){\vector(-1,0){30}}
\put(22,170){\vector(0,-1){20}}
\put(77,170){\vector(0,-1){20}}
\put(160,170){\vector(0,-1){20}}
\put(270,170){\vector(0,-1){20}}
\put(0,100){\framebox(100,50){{\shortstack{Instanton expansion\\
				of $\cnt$ prepotential \\
				from microscopic \\Lagrangian}}}}
\put(110,100){\framebox(100,50){{\shortstack{Instanton expansion\\
				of $\cnt$ prepotential\\
				from SW theory}}}}
\put(220,100){\framebox(100,50){{\shortstack{Instanton expansion\\
				of $\cnt$ prepotential\\
				from DV \\matrix model}}}}
\put(150,100){\line(0,-1){20}}
\put(150,80){\line(-1,0){100}}
\put(50,80){\vector(0,1){20}}
\put(50,60) {\makebox(100,20){{test of SW theory}}}
\put(270,100){\line(0,-1){20}}
\put(270,80){\line(-1,0){100}}
\put(170,80){\vector(0,1){20}}
\put(170,60) {\makebox(100,20){{test of matrix model}}}
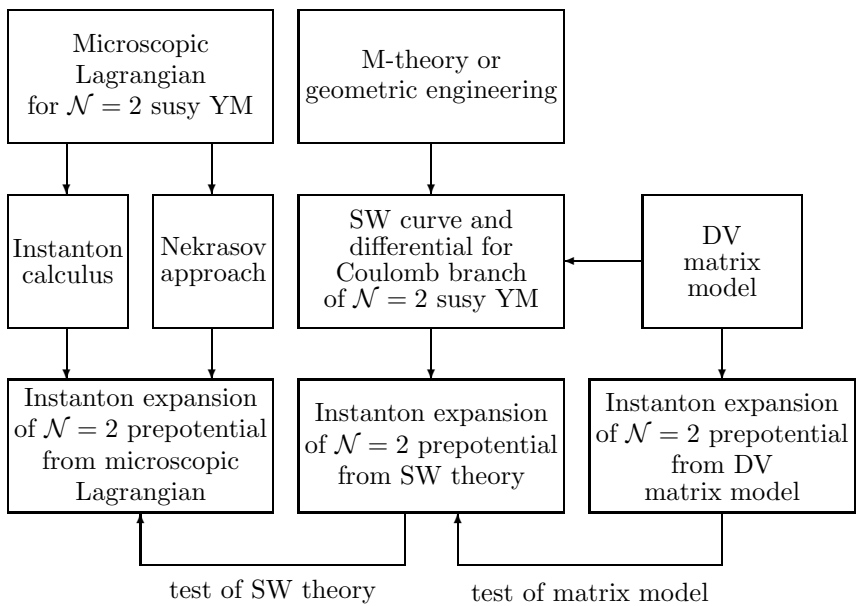
\put(0,30) {\makebox(300,20){{\bf Figure 2: Computing the instanton expansion}}}
\end{picture}

\renewcommand{\theequation}{3.\arabic{equation}}
\setcounter{equation}{0}
\section{The SW curve from IIA string theory/M theory}

Klemm, {\it et.~al.},\cite{003a} demonstrated that SW theory
could be derived from string theory, using a technique called 
geometric engineering.
Subsequently, Witten\cite{003} gave a systematic method 
to find SW curves by lifting 10-dimensional IIA string theory 
to 11-d M-theory.  
In IIA language, this involve configurations of two (or more) 
parallel NS5 branes spanned by a
number of D4 branes between each NS5 neighboring pair.  
For pure $\cnt$ SU($N$) gauge theory (with no hypermultiplets),
one has two parallel NS5 branes, connected by $N$ D4 branes,
as shown in fig.~3. 

\begin{picture}(330,220)(0,0)

\put(100,50){\line(0,1){150}} \put(220,50){\line(0,1){150}}

\put(240,50){\vector(1,0){30}} \put(274,48){$x_6$}


\put(70,120){\vector(0,1){30}} \put(69,153){$v$}
\put(42,110){$(x_4+ix_5)$}

\put(100,80){\line(1,0){9}} \put(119,80){\line(1,0){9}}
\put(138,80){\line(1,0){9}} \put(157,80){\line(1,0){9}}
\put(176,80){\line(1,0){9}} \put(195,80){\line(1,0){9}}
\put(211,80){\line(1,0){9}}

\put(100,100){\line(1,0){9}} \put(119,100){\line(1,0){9}}
\put(138,100){\line(1,0){9}} \put(157,100){\line(1,0){9}}
\put(176,100){\line(1,0){9}} \put(195,100){\line(1,0){9}}
\put(211,100){\line(1,0){9}}

\put(100,180){\line(1,0){9}} \put(119,180){\line(1,0){9}}
\put(138,180){\line(1,0){9}} \put(157,180){\line(1,0){9}}
\put(176,180){\line(1,0){9}} \put(195,180){\line(1,0){9}}
\put(211,180){\line(1,0){9}}

\put(155,120){$\cdot$} \put(155,125){$\cdot$}
\put(155,130){$\cdot$} \put(155,115){$\cdot$}

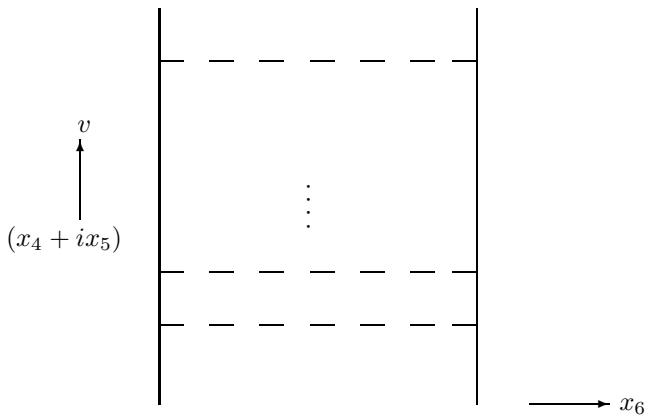
\put(0,20){\makebox(300,20)[b]{\bf {Figure 3: IIA brane configuration
for $\cnt$ SU($N$) gauge theory}}}

\end{picture}

In the lift to 11-d one can visualize the IIA brane picture as a Riemann
surface, with (fattened) lines for the D4 branes becoming
branch-cuts which connect two Riemann sheets, {\it i.e.}, the two
NS5 branes.
This gives the hyperelliptic curve
\beq
y^2 = \prod^N_{i=1} (x-e_i)^2 - 4\Lambda^{2N} \; .
 \eeq

More interesting are the theories yielding {\it non}-hyperelliptic 
curves, particularly since, for a long time,
string theory was the only method available to obtain such
curves. 
For example, the curve for
$ {\rm SU}(N_1) \times {\rm SU}(N_2)$
gauge theory, with hypermultiplets in bifundamental  representations
$(N_1,\bar{N}_2)$ and $(\bar{N}_1,N_2)$,
is obtained from 3 parallel NS5 branes, with $N_1$ D4-branes
between the left and middle NS5 branes, and $N_2$ D4-branes
between the middle and right NS5's. 
Upon lift to M-theory, this configuration yields a 
3-sheeted surface of the form\cite{003a,003}
 \beq
\label{eq:nonhyper}
 y^3 + A(x)y^2 + B(x)y + C(x) =0
 \eeq
where the polynomials $A(x)$, $B(x)$, and $C(x)$
depend on the details of the theory.
Even more interesting is the case when an 
orientifold $O6$ plane is placed on the middle
NS5 brane\cite{004}, as shown in fig.~4. 
The presence of the orientifold plane implies a reflection symmetry 
on the brane-picture, and an involution imposed on 
a curve of type (\ref{eq:nonhyper}).
An $O6^+$ plane corresponds to ${\rm SU}(N) + \Ysymm$ gauge theory, 
while an $O6^-$ plane yields ${\rm SU}(N) + \Yasymm$.

\begin{picture}(330,220)(80,00)

\put(100,50){\line(0,1){150}} \put(220,50){\line(0,1){150}}
\put(340,50){\line(0,1){150}}

\put(350,50){\vector(1,0){30}} \put(384,48){$x_6$}


\put(70,120){\vector(0,1){30}} \put(69,153){$v$}

\put(216,124){$\otimes$} \put(225,124){$O6$}

\put(100,80){\line(1,0){9}} \put(119,80){\line(1,0){9}}
\put(138,80){\line(1,0){9}} \put(157,80){\line(1,0){9}}
\put(176,80){\line(1,0){9}} \put(195,80){\line(1,0){9}}
\put(211,80){\line(1,0){9}}

\put(100,100){\line(1,0){9}} \put(119,100){\line(1,0){9}}
\put(138,100){\line(1,0){9}} \put(157,100){\line(1,0){9}}
\put(176,100){\line(1,0){9}} \put(195,100){\line(1,0){9}}
\put(211,100){\line(1,0){9}}

\put(100,161){\line(1,0){9}} \put(119,161){\line(1,0){9}}
\put(138,161){\line(1,0){9}} \put(157,161){\line(1,0){9}}
\put(176,161){\line(1,0){9}} \put(195,161){\line(1,0){9}}
\put(211,161){\line(1,0){9}}

\put(155,120){$\cdot$} \put(155,125){$\cdot$}
\put(155,130){$\cdot$} \put(155,115){$\cdot$}

\put(220,90){\line(1,0){9}} \put(239,90){\line(1,0){9}}
\put(258,90){\line(1,0){9}} \put(277,90){\line(1,0){9}}
\put(296,90){\line(1,0){9}} \put(315,90){\line(1,0){9}}
\put(331,90){\line(1,0){9}}

\put(220,152){\line(1,0){9}} \put(239,152){\line(1,0){9}}
\put(258,152){\line(1,0){9}} \put(277,152){\line(1,0){9}}
\put(296,152){\line(1,0){9}} \put(315,152){\line(1,0){9}}
\put(331,152){\line(1,0){9}}

\put(220,170){\line(1,0){9}} \put(239,170){\line(1,0){9}}
\put(258,170){\line(1,0){9}} \put(277,170){\line(1,0){9}}
\put(296,170){\line(1,0){9}} \put(315,170){\line(1,0){9}}
\put(331,170){\line(1,0){9}}

\put(275,120){$\cdot$} \put(275,125){$\cdot$}
\put(275,130){$\cdot$} \put(275,135){$\cdot$}
\put(80,20){\makebox(300,20)[b]{\bf {Figure 4: IIA brane configuration
for $\cnt$ ${\rm SU}(N) + \Ysymm / \Yasymm$ }}}
\end{picture}

One of the significant contributions of the Brandeis group is the
development of methods for finding the instanton expansion for
 non-hyperelliptic curves such as (\ref{eq:nonhyper}), 
and others\cite{005}.  In order to
 compute the instanton expansion for the prepotential one needs to
 compute
 \beq
 a_i  \sim  \oint_{A_i} \: \frac{x\,dy}{y} \; ,
 \qquad\qquad 
 (a_D)_i  \sim  \oint_{B_i} \: \frac{x\,dy}{y}
 \eeq
in some approximation scheme since an exact solution seems
inaccessible.  
The method developed by our group was hyperelliptic
perturbation theory\cite{005}, {\it i.e.}, a systematic expansion
of the non-hyperelliptic curve about a fiducial hyperelliptic
curve.

The treatment of SU$(N) +  \Ysymm$ and SU$(N) + \Yasymm$ 
gauge theories typify the
method, but a number of other examples including other groups were
also considered.  Explicit calculations enabled us to obtain
${\cF}_{\rm 1-inst}$ from our methods.  More recent 
calculations\cite{006,007} involving the renormalization group (RG) have
simplified the method, and allow an easier access to two or more
instanton contributions.  In {\it every} case, the results from
our hyperelliptic expansion agree with those of microscopic
calculations, when available.  The most systematic of these are by
Nekrasov and others\cite{008}, 
and recently by Mari\~no and Wyllard\cite{009}.  
This agreement supports the M-theory approach
to SW theory, as well as the validity of our approximation
methods.

\renewcommand{\theequation}{4.\arabic{equation}}
\setcounter{equation}{0}

\section{The Dijkgraaf--Vafa matrix model approach}

A major advance in our understanding of susy gauge theory
was provided by Dijkgraaf, Vafa, and collaborators\cite{002}.
They showed that a suitably formulated matrix model will describe
the low-energy physics of $\cno$ or $\cnt$ susy gauge
theories. 
This is proved by showing that, in the computation of the
effective superpotential $W_{\rm eff}$ and the period matrix
$\tau_{ij}$ of the gauge theory, 
the space-time part of Feynman integrals cancels,
leaving a zero-dimensional theory, {\it i.e.}, a matrix model\cite{010}.
An alternate proof\cite{011} shows that the generalized Konishi
anomaly equations of the gauge theory have a  form identical to those 
of the resolvent operator of the matrix model.

We will describe the application of the DV model to SW theory. To
be specific, consider the matrix model appropriate to $\cnt$
U$(N)$ theory with $N_f$ hypermultiplets in the fundamental ($\Yfund$)
representation.  
Consider $M\times M$ matrices $\Phi$ together
with $M$-vectors $Q^I, \; \tilde{Q}_I \; (I = 1 \; {\rm to} \; N_f)$.  
One wishes to compute the partition function\cite{002}
\beq
\label{eq:part}
Z = \frac{1}{{\rm vol} \: G}\: \int \: \dd\Phi \; \dd Q^I \; \dd\tilde{Q}_I
\exp \left[ - \frac{1}{g_s} \, W (\Phi , Q, \tilde{Q} ) \right]\;  \eeq 
with $G$ the unbroken matrix gauge group.
The choice of $W(\Phi , Q, \tilde{Q})$ determines the
physics.  For our example
\beq
\label{eq:action}
W = W_0 (\Phi ) + W_{\rm matter}(\Phi , Q, \tilde{Q} )
 \eeq
 where
 \vspace{-.1in}
\beq
W^\prime_0 (x) = \alpha \: \prod^N_{i=1} (x - e_i )
 \eeq
 breaks $\cnt$ susy to  $\cno$.  One takes $\alpha\rightarrow 0$
 at the end of the calculation to recover $\cnt$  physics
 from the $\alpha$-independent quantities $W_{\rm eff}$ and $\tau_{ij}$.
 The matter interaction in (\ref{eq:action}) is taken to be of the same form
 as the analogous superpotential in the underlying gauge theory.
 For the theory we are considering,\cite{012,013}
 \beq
W_{\rm matter} = \sum^M_{a,b} \, \sum^{N_f}_{I=1}\:
\tilde{Q}_{aI} \, \Phi^{a}{}_{b} \, Q^{bI}  .
 \eeq

To evaluate the partition function (\ref{eq:part}),
one expands the action about a stationary point\cite{002}
\beq
 \Phi = \Phi_0 \; , \qquad\; Q=\tilde{Q} = 0
 \eeq
with fluctuations
 \beq
 \Phi=  \Phi_0 + \Psi = 
 {\rm diag} \: (e_i \: \textstyle{1\!\!1_{M_i} }) + (\Psi_{ij})
 \eeq
where $\Psi_{ij}$ is an $M_i \times M_j$ matrix, and  
\beq
\sum_{i=1}^N \: M_i = M \; , \qquad\qquad  G = \prod^N_{i=1} \: U(M_i)\; .
 \eeq
One makes the gauge choice $\Psi_{ij}=0$ for $i \neq j$,
which requires the introduction of ghost matrix fields\cite{017}.  
One obtains the Feynman rules from the expansion of the action,
as well as for the ghost contributions,  
from which one can do a  perturbative evaluation of the
partition function. 

In the limit  $M_i \gg 1$ for all $i$, 
with
 \beq
 S_i \equiv g_s \; M_i = {\rm finite},
 \eeq
one may express the partition integral in a topological expansion\cite{002},
 \beq
 Z =  \exp \: F(e,S) =
 \exp \: \sum_\chi \: \frac{1}{g_s^{~\chi}} \: F_\chi (e,S)
 \eeq
 where $F(e,S)$ is the free-energy and $\chi = 2-2g-h$ is the
 Euler number for a two-dimensional surface with $g$ handles and
 $h$ holes.  
The leading terms are
 \beq
 F(e,S) = 
\frac{1}{g_s^2} F_{\rm sphere} (e,S) +  
\frac{1}{g_s} F_{\rm disk} (e,S) + \ldots
 \eeq
 where the sphere contribution has $g=h=0$, while
 the disk contribution has $g =0$, $h =1$,  
the latter arising from diagrams  with $Q$ or $\tilde{Q}$ running
along the boundary.  
(The $h \geq 2$ diagrams are suppressed relative to 
$\chi = 1$ for the limit $g_s \rightarrow 0$, 
$M_i \rightarrow \infty$, with $g_s M_i = S_i$ fixed.)

On the gauge theory side, the gauge symmetry is broken to
 \beq
 U(N) \rightarrow \prod_i \: U(N_i)\; \qquad  \sum_i N_i = N
 \eeq
It is important to emphasize that $M_i \rightarrow \infty$ on the
matrix-model side, while $N_i$ remains finite on the gauge theory side. 
In order to study a generic point on the Coulomb branch 
of the $\cnt$  SW theory, one takes $N_i = 1$ for all $i$, 
that is,
\beq
U(N) \rightarrow [U(1)]^N \; .
 \eeq
The effective superpotential of the gauge theory 
is then given by\cite{002,012,013}
 \beq
 W_{\rm eff} (e,S) =  - \sum^N_{i=1} \:
 \frac{\partial \, F_{\rm sphere}(e,S) }{\partial S_i} \: - 
F_{\rm disk}(e,S) 
  +  2\pi i \: \tau_0 \sum^N_{i=1} \: S_i
 \eeq
 where $\tau_0 = \tau (\Lambda_0 )$ is the gauge coupling of U$(N)$
at scale $\Lambda_0$.
To proceed, one finds the extremum 
\beq
\frac{\partial W_{\rm eff}}{\partial S_i}\Bigg|_{\vev{S_j}} = 0
 \eeq
which defines $\vev{S_i}$.  

The period matrix is then obtained 
from {\it only} the sphere contribution 
to the free-energy 
\beq
\label{eq:taue}
\tau_{ij} (e) = \frac{1}{2\pi i} \; \frac{\partial^2 F_{\rm
sphere}(e,S)}{\partial S_i\partial S_j}\Bigg|_{\vev{S_i}}
 \eeq
One lets $\alpha \rightarrow 0$ to obtain $\cnt$ results.  
(Actually, however,  eq.~(\ref{eq:taue}) is independent of $\alpha$.)

While eq.~(\ref{eq:taue}) gives the matrix of U$(1)^N$
couplings as a function of the $\{ e_i \}$, 
the period matrix (\ref{eq:taudef}) from SW theory,
\beq
\tau_{ij} (a) = \frac{\partial^2 {\cF} (a)}{\partial a_i\partial a_j},
 \eeq
is expressed in terms of {\it different} parameters on
moduli space.  
The $\{ a_i \}$ are physical order parameters, 
while $\{ e_i \}$ are ``bare" order parameters, 
so one needs a relation between $a_i$ and  $e_i$.
One can show that the desired relation is\cite{012}
 \beq 
\label{eq:tadpole}
a_i  =  e_i + 
\left[ \sum^N_{j=1} \frac{\partial}{\partial S_j} \:
g_s \langle \tr \: \Psi_{ii} \rangle _{\rm sphere} + \langle \tr
\: \Psi_{ii} \rangle_{\rm disk} \right]_{\vev{S}}
 \eeq
where $\tr$ is the trace over the $i^{\rm th}$ diagonal block 
in $\Psi$ only.  
The computation of (\ref{eq:tadpole}) involves the
calculation of tadpole diagrams with external $\Psi_{ii}$ legs
in the matrix model.
Using eq.~(\ref{eq:tadpole}), one may re-express (\ref{eq:taue}) 
as a function of $a_i$.
Finally, integration of $\tau_{ij} (a)$
yields the $\cnt$ prepotential
$\cF(a)$.
The procedure just outlined for calculating $\cF(a)$ is
shown schematically in fig.~5.

\begin{picture}(320,400)(0,50)

\put(100,400){\framebox(100,30){{\shortstack{DV Matrix model}}}}
\put(150,400){\vector(0,-1){20}}
\put(70,330){\framebox(160,50){{\shortstack{Topological expansion \\
of free energy\\
$F(e,S) = g_s^{-2} F_{\rm sph} + g_s^{-1} F_{\rm disk} + \cdots$ 
}}}}
\put(150,330){\vector(0,-1){20}}
\put(230,330){\vector(1,-1){20}}
\put(0,260){\framebox(80,50){{\shortstack{Tadpole \\
	calculation \\
		relates $a_i$ to $e_i$}}}}
\put(40,260){\vector(0,-1){20}}
\put(90,260){\framebox(120,50){{\shortstack{
$\tau_{ij}(e) = 
\frac{1}{ 2 \pi i} \frac{\partial^2 F_{\rm sph} (e,S)}
     {  \partial S_i \partial S_j }  \bigg|_{\vev{S_i}}  $ }}}}
\put(220,260){\framebox(80,50){{\shortstack{$W_{\rm eff} (e,S)$ \\
from $F(e,S)$}}}}
\put(140,260){\vector(-1,-1){20}}
\put(260,260){\vector(0,-1){20}}
\put(0,200){\framebox(120,40){{\shortstack{
$\tau_{ij}(a)  = \frac{\partial^2 \cF(a)}{ \partial a_i \partial a_j } $}}}}
\put(220,200){\framebox(80,40){{\shortstack{
$\frac{\partial W_{\rm eff}}{ \partial S_i} = 0$  
\\ $\implies \vev{S_i}$    }}}}
\put(220,240){\vector(-1,1){20}}
\put(60,200){\vector(0,-1){20}}
\put(0,150){\framebox(120,30){{\shortstack{
$\cF(a)$ \\
($\cnt$ prepotential) }}}}
\put(260,200){\vector(-1,-1){20}}
\put(180,150){\framebox(120,30){{\shortstack{$W_{\rm eff} (\vev{S_i})$
\\ ($\cno$ superpotential) }}}}
\put(60,150){\vector(0,-1){20}}
\put(60,130) {\makebox(50,20){compare}}
\put(60,130){\vector(0,1){20}}
\put(0,100){\framebox(120,30){{\shortstack{
$\cnt$ prepotential \\ from SW theory }}}}

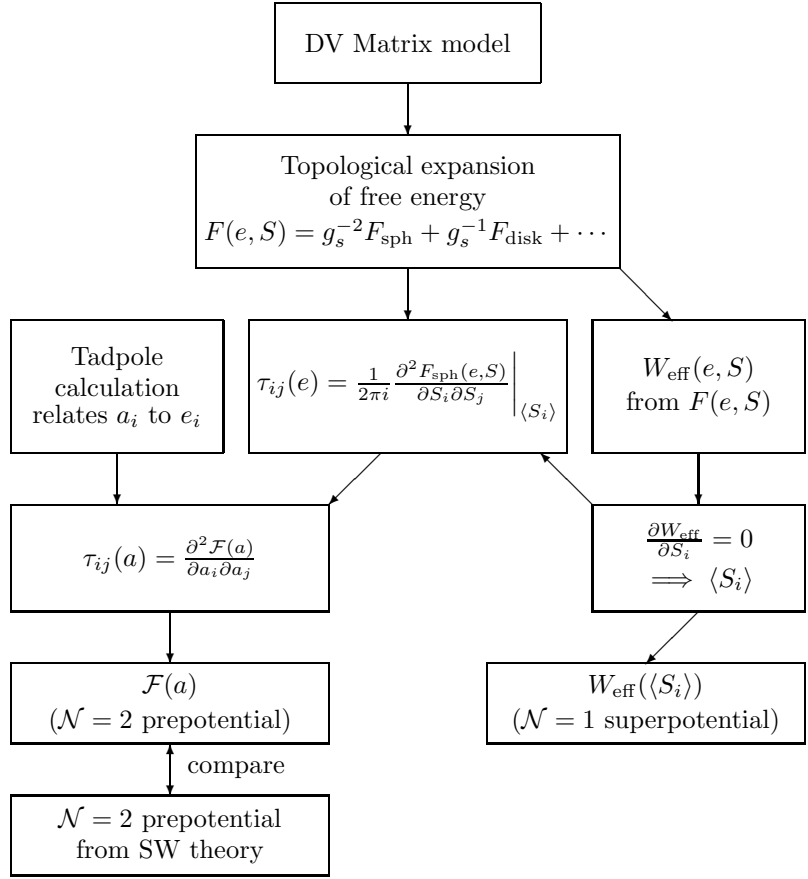
\put(0,70) {\makebox(300,20){{\bf Figure 5:  Matrix model flow chart}}}
\end{picture}

In practice, one computes the matrix-model
quantities to a certain power in $\Lambda$, 
which increases with the number of loops in the matrix model, 
but corresponds to the instanton expansion in gauge theory.
That is, non-perturbative information in the gauge theory 
is obtained from perturbative calculations in the matrix model!  
In these calculations $(n+1)$-loop perturbation theory in the matrix model 
corresponds to the $n$-instanton term of the prepotential of 
the gauge theory\cite{012}.
Recent work\cite{007}  using the renormalization group 
has improved the situation, so that only 
$n$-loops in the matrix model theory are required to obtain $n$-instanton
accuracy in the gauge theory.
Another improvement\cite{014} allows one to obtain the
one-instanton prepotential for U$(N) + N_f \Yfund$
using only the contribution from $F_{\rm sphere}$.
In {\it every} case, our matrix model calculations 
agree with results from
``conventional" SW theory, and microscopic calculations.

{}From the matrix model approach, 
one can also directly derive\cite{017,012} the SW curve and differential
from matrix-model resolvent equations.
Particularly noteworthy is the treatment of 
{\it non}-hyperelliptic curves within the context of matrix
models\cite{015}. 
It is possible to derive the correct SW curve
for U$(N) + \Ysymm$  and U$(N) + \Yasymm$ gauge theories, 
as well as the one-instanton contribution to the prepotential. 
There are subtle points\cite{015}  in constructing $\tau_{ij}$ for 
these two theories, and additional
subtle issues\cite{015,016} in choosing the correct 
matrix-model vacuum state for U$(N) +\Yasymm$.
When these issues are dealt with correctly, 
one obtains agreement with our previous results from
hyperelliptic perturbation theory, as well as microscopic
calculations of Nekrosov and others,\cite{008} 
and Mari\~{n}o and Wyllard\cite{009}. 

\section{Concluding remarks}

Very often a single problem plays a central role in testing a wide
variety of methods.  Seiberg--Witten theory may not be the
hydrogen atom of strongly interacting gauge theories, but it does
seem to have a privileged position in regard to $\cnt$ susy gauge
theories.  As our initial road-map indicated, SW theory can be
treated by rather diverse techniques,  among them 
IIA string theory/M theory, 
the DV matrix model, microscopic instanton calculations,
and integrable systems\cite{018,019}.  We have surveyed just two
of these approaches due to time limitations, but the subject is
vast and still developing in the other areas as well.

HJS wishes to thank the organizers of the Deserfest for the
opportunity to honor Stanley with this survey of topics which have
occupied us for a long time.

\end{document}